\documentclass[aps,twocolumn,preprintnumbers,amsmath,amssymb,superscriptaddress,floatfix,nofootinbib]{revtex4}
\usepackage{lipsum} \usepackage{graphicx} 
\usepackage{epsfig} \usepackage{epstopdf} \usepackage{hyperref}
\usepackage{amsmath} \usepackage{amsfonts} \usepackage{amssymb}

\DeclareMathSizes{12}{12}{12}{12}
\usepackage[section]{placeins}

\begin{document}

\title{$\Omega_c$ excited states within a ${\rm SU(6)}_{\rm lsf}\times$HQSS model}

\author{J.~Nieves}
\affiliation{Instituto~de~F\'{\i}sica~Corpuscular~(centro~mixto~CSIC-UV),
  Institutos~de~Investigaci\'on~de~Paterna, Aptdo.~22085,~46071,~Valencia,
  Spain}
  
\author{R.~Pavao}
\affiliation{Instituto~de~F\'{\i}sica~Corpuscular~(centro~mixto~CSIC-UV),
  Institutos~de~Investigaci\'on~de~Paterna, Aptdo.~22085,~46071,~Valencia,
  Spain}

  \author{L.~Tolos}
  \affiliation{Institut f\"ur Theoretische Physik, University of Frankfurt, Max-von-Laue-Str. 1, 60438 Frankfurt am Main, Germany}
  \affiliation{Frankfurt Institute for Advanced Studies, University of Frankfurt, Ruth-Moufang-Str. 1,
60438 Frankfurt am Main, Germany}
  \affiliation{Institute of Space Sciences (ICE, CSIC), Campus UAB, Carrer de Can Magrans, 08193, Barcelona, Spain}
\affiliation{Institut d'Estudis Espacials de Catalunya (IEEC), 08034 Barcelona, Spain}

  \date{\today}

\begin{abstract}
We have reviewed the renormalization procedure  used in the unitarized coupled-channel model
of Phys.\ Rev.\ D {\bf 85}  114032 (2012), and its impact in the
$C=1$, $S=-2$, and $I=0$ sector, where five $\Omega_c^{(*)}$ states have been recently
observed by the LHCb Collaboration. The meson-baryon interactions used
in the model are consistent with both chiral and heavy-quark
spin symmetries, and lead to a successful description of the
observed lowest-lying
odd parity resonances $\Lambda_c(2595)$ and $\Lambda_c(2625)$, and
$\Lambda_b(5912)$ and $\Lambda_b(5920)$ resonances. We show that some (probably at least three) of the states observed by
LHCb  will also have odd parity and $J=1/2$ or
$J=3/2$, belonging two of them to the same  ${\rm SU(6)}_{\rm light-spin-flavor}\times$HQSS multiplets as the latter charmed and beauty $\Lambda$ baryons.
  \end{abstract}
  
\maketitle


\section{Introduction}
\label{intro}
The LHCb Collaboration \cite{Aaij:2017nav} has recently reported the
existence of five $\Omega_c$ states, analyzing the $\Xi_c^+ K^-$
spectrum in $pp$ collisions, with masses ranging between 3 and 3.1 GeV. These results have renewed the interest in baryon spectroscopy, with the long-standing question whether these states can be accommodated within the quark model picture and/or qualify better as being dynamically generated via hadron-hadron scattering processes.

Earlier predictions for such states have been reported within
conventional quark models
\cite{Chiladze:1997ev,Ebert:2007nw,Roberts:2007ni,Garcilazo:2007eh,Migura:2006ep,Ebert:2011kk,Valcarce:2008dr,Shah:2016nxi,Vijande:2012mk,Yoshida:2015tia,Chen:2015kpa,Chen:2016phw}.
The experimental discovery of the five $\Omega_c$ states has triggered
a large activity in the field, and thus  some quark models have been revisited in view of the new results \cite{Karliner:2017kfm,Wang:2017hej,Almasi:2017bhq,Chen:2017gnu,Zhao:2017fov,Cheng:2017ove,Wang:2017vnc}, suggestions as pentaquarks have been advocated \cite{Yang:2017rpg,Huang:2017dwn,Kim:2017jpx,An:2017lwg,Anisovich:2017aqa}, models based on QCD sum-rules have been put to test \cite{Agaev:2017jyt,Agaev:2017lip,Chen:2017sci,Wang:2017zjw,Aliev:2017led,Mao:2017wbz,Agaev:2017ywp}, or quark-soliton models have been employed \cite{Wang:2017kfr}. Also, Lattice QCD has reported results on the spectroscopy of $\Omega_c$ states \cite{Padmanath:2017lng}.

Within molecular models, there have been previous predictions on
$\Omega_c$ states
\cite{JimenezTejero:2009vq,Hofmann:2005sw,GarciaRecio:2008dp, Romanets:2012hm}.  
In Ref.~\cite{Hofmann:2005sw} several resonant states were obtained
with masses much below 3 GeV, by employing a zero-range exchange of
vector mesons as the bare interaction for the $s$-wave baryon-meson
scattering. Similar qualitative results were obtained in
Ref.~\cite{JimenezTejero:2009vq}, where finite range effects were
considered.  Lately the work of  
Ref.~\cite{Montana:2017kjw} has revisited Ref.~\cite{Hofmann:2005sw},  finding that, after modifying the
regularization scheme with physically motivated parameters, two
$\Omega_c$ resonant states were generated at 3050 MeV and 3090 MeV
with spin-parity $J^P=1/2^-$, reproducing the masses and widths of two
of the experimental states. More recently, the $\Omega_c$ states have
been also investigated using an extended local hidden gauge
approach~\cite{Debastiani:2017ewu}. Within this scheme,  low-lying $1/2^+$ and $3/2^+$ baryons, as well as
pseudoscalar and vector mesons,  are considered to construct the
baryon-meson coupled channel space. In this manner, two $\Omega_c$ states of $J^P=1/2^-$ and one $\Omega_c^*$ $J^P=3/2^-$ can be identified, the first two in good agreement with the results of \cite{Montana:2017kjw} and the third one fairly well.

The use of the hidden-gauge formalism allows for the preservation of
heavy-quark spin symmetry (HQSS), which is a proper QCD symmetry that
appears when the quark masses, such as that of the  charm quark, become larger
than the typical confinement scale. Aiming to incorporate explicitly
HQSS, a scheme was developed in
Refs.~\cite{GarciaRecio:2008dp,Gamermann:2010zz,Romanets:2012hm,GarciaRecio:2012db,Garcia-Recio:2013gaa}
that implements a consistent ${\rm SU(6)}_{\rm lsf} \times {\rm
  SU(2)}_{\rm HQSS}$ extension of the Weinberg-Tomozawa (WT) $\pi N$
interaction, where ``lsf'' stands for light quark-spin-flavor symmetry, respectively. Indeed, the works of
Refs.~\cite{GarciaRecio:2008dp, Romanets:2012hm} are the first
meson-baryon molecular studies, fully consistent with HQSS, of the
well-established odd-parity $\Lambda_c(2595)$ [$J=1/2$] and
$\Lambda_c(2625)$ [$J=3/2$] resonances.

Within this scheme in the $J=1/2$ sector, one finds a pole
structure that mimics the well-known two-pole pattern of the
$\Lambda(1405)$~\cite{GarciaRecio:2002td, Ramos:2002xh,
  GarciaRecio:2003ks,Oller:2000fj,Jido:2003cb}. Thus, in the region of 2595 MeV, two states are
dynamically generated. The first one, identified with the
$\Lambda_c(2595)$ resonance, is narrow and strongly couples to the
$ND$ and $ND^*$ channels, with a negligible coupling to the open
$\Sigma_c\pi$ channel.  The second state is quite broad and it has a
sizable coupling to this latter channel. On the other hand, the
$J^P=(3/2)^-$ state is generated mainly by the $(ND^*, \Sigma_c^*\pi
)$ coupled-channel dynamics, and it would be the charm counterpart of
the $\Lambda(1520)$. Similar results are also obtained in the extension 
of the local hidden gauge approach of Ref.~\cite{Liang:2014kra}.
The same scheme also dynamically generates the
$\Lambda_b(5912)$ and $\Lambda_b(5920)$ narrow resonances, discovered
by LHCb in 2012 \cite{Aaij:2012da}, which turn out to be HQSS
partners, naturally explaining in this way their
approximate mass degeneracy~\cite{GarciaRecio:2012db}. Moreover, the $\Lambda_b(5920)$ resonance
turns out to be the bottom version of the $\Lambda_c(2625)$ one, while
the $\Lambda_b(5912)$ would not be the counterpart of the
$\Lambda_c(2595)$ resonance, but it would be of the second charmed
state that appears around 2595 MeV, and that gives rise to the
two-pole structure mentioned above~\cite{GarciaRecio:2012db}.

In Ref.~\cite{Romanets:2012hm} five $\Omega_c$ states were found, three $J=1/2$ and the two $J=3/2$ bound states, the positions being shown in Table VI of that
reference or in Table~\ref{tab:table1} in the present work.
These states come from the most attractive ${\rm SU(6)}_{\rm
  lsf}\times$HQSS representations.   Attending to the breaking pattern
of the spin-flavor SU(8) symmetry discussed in
Ref.~\cite{Romanets:2012hm}, the two lowest-lying $\Omega_c$ and
$\Omega^*_c$ states
({\bf a} and {\bf b}) and the $\Lambda_c(2595)$ would be members of the
same {\bf 21} ${\rm SU(6)}_{\rm lsf}$ multiplet, while both, the third
$\Omega_c$ ({\bf c}) and
the  $\Lambda_c(2625)$ resonances would be in the {\bf 15} ${\rm SU(6)}_{\rm lsf}-$
irreducible representation. Finally, the two heaviest
$\Omega_c$ and $\Omega^*_c$ states ({\bf d} and {\bf e}) reported in
\cite{Romanets:2012hm} would not be 
directly related to the  $\Lambda_c(2595)$ and
$\Lambda_c(2625)$ resonances, since they would stem originally from a
different SU(8) representation.  These five odd-parity
$\Omega_c, \Omega^*_c$ states, coming from the most attractive ${\rm
  SU(6)}_{\rm lsf}\times$ HQSS representations, have 
masses below 2.98 GeV, and cannot be easily identified with any of the LHCb
resonances, located all of them above 3 GeV. Predicted masses,
however,  depend not only on the baryon-meson interactions, but also
on the adopted renormalization scheme (RS). In this work we review the RS used in  \cite{Romanets:2012hm}, and its impact in the generation of the $\Omega_c^{(*)}$ states. We show how the pole positions can be moved up by implementing a different RS, making then feasible the
identification of at least three states with the observed $\Omega_c^{(*)}$ states by LHCb.

The paper is organized as follows. In Section \ref{formalism} we present the ${\rm SU(6)}_{\rm lsf} \times {\rm SU(2)}_{\rm HQSS}$
extension of the WT interaction, while in Section \ref{sec:res} we show our results for the $\Omega_c^{(*)}$ states and the possible identification of three of them with the experimental ones. Finally, in Section \ref{conc} we present our conclusions.

\begin{table}[t]
  \caption{ $\Omega_c$ an $\Omega_c^*$ states, reported in
    Ref.~\cite{Romanets:2012hm}, coming from the most attractive 
${\rm SU(6)}_{\rm lsf} \times$ HQSS representations. We label those
    states from {\bf a} to {\bf e}, according to their position in energy.}
  \label{tab:table1}
  \begin{tabular}{c|c|c|c|}
      Name & $M_R$ (MeV) & $\Gamma_R$ (MeV) & $J$  \\
    \hline
    {\bf a} & 2810.9 & 0 & 1/2  \\
    \hline
    {\bf b} & 2814.3 & 0 & 3/2  \\
    \hline
    {\bf c} & 2884.5 & 0 & 1/2  \\
    \hline
    {\bf d} & 2941.6 & 0 & 1/2  \\
    \hline
   {\bf e} & 2980.0 & 0 & 3/2  \\
  \end{tabular}
\end{table}

\section{Formalism}
\label{formalism}

We will consider the sector with charm
$C=1$, strangeness $S=-2$ and isospin $I=0$ quantum numbers, where the
$\Omega_c^{(*)}$ excited states are located by revising the results in
Ref.~\cite{Romanets:2012hm}.

The building-blocks in the $C=1$ sector are the pseudoscalar ($D_s, D,
K, \pi,\eta, {\bar K}, {\bar D}, {\bar D}_s$) and vector ($D_s^*,
D^*,K^*, \rho,\omega, {\bar K}^{*}, {\bar D}^*, {\bar D}_s^*, \phi$)
mesons, the spin--$1/2$ octet and the spin--$3/2$ decuplet of
low-lying light baryons, in addition to the spin-1/2 ($\Lambda_c$,
$\Sigma_c$, $\Xi_c$, $\Xi'_c$, $\Omega_c$), and spin-3/2
($\Sigma^*_c$, $\Xi^*_c$, $\Omega_c^*$) charmed
baryons~\cite{Romanets:2012hm, Garcia-Recio:2013gaa}. All baryon-meson pairs with
$(C=1, S=-2, I=0)$ quantum numbers span the coupled-channel space for
a given total angular momentum ($J$). The $s$-wave tree level
amplitudes between two channels are given by the ${\rm SU(6)}_{\rm lsf}$ $\times$ HQSS
WT  kernel
\begin{equation}
\label{eq:WT}
V_{ij}^J(s) = D_{ij}^J \frac{2 \sqrt{s}-M_i-M_j}{4 f_i f_j} \sqrt{\frac{E_i+M_i}{2 M_i}} \sqrt{\frac{E_j+M_j}{2M_j}},
\end{equation}
with $M_i$ and $m_i$, the masses of the baryon and meson in the $i$
channel, respectively, and $E_i$ the center-of-mass
energy of the baryon in the same channel,
\begin{equation}
E_i=\frac{s-m_i^2+M_i^2}{2 \sqrt{s}}.
\end{equation}
The hadron masses and meson decay constants, $f_i$,  have been taken from
Ref.~\cite{Romanets:2012hm}. The $D_{ij}^J$ matrices are determined by
the underlying ${\rm SU(6)}_{\rm lsf} \times$ HQSS group
structure of the interaction. Tables for all of them
can be found in the Appendix B of Ref.~\cite{Romanets:2012hm}.

We use the matrix $V_{ij}^J$ as potential to solve the Bethe-Salpeter
equation (BSE), which leads to a $T$-matrix of the form
\begin{equation}
\label{eq:LS}
T^J(s)=\frac{1}{1-V^J(s) G^J(s)} V^J(s),
\end{equation}
satisfying exact unitarity in coupled channels. In the above equation,
$G^J(s)$ is a diagonal matrix that contains the loop functions
corresponding to the particles of the different channels being considered.

The two-body loop function is given by
\begin{equation}
\label{eq:normloop}
G_i(s)=i 2M_i  \int \frac{d^4 q}{(2 \pi)^4} \frac{1}{q^2-m_i^2+i\epsilon} \frac{1}{(P-q)^2-M_i^2+i\epsilon},
\end{equation}
with $P$ the total momentum of the system such that $P^2=s$. We omit
the index $J$ from here on for simplicity. The bare loop function is
logarithmically ultraviolet (UV) divergent and needs to be
renormalized. This can be done by one-subtraction
\begin{equation}
G_i(s)=\overline{G}_i(s)+G_i(s_{i+}) ,
\label{eq:div}
\end{equation}
with the finite part of the loop function, $\overline{G}_i(s)$,
given in Ref.~\cite{Nieves:2001wt},
\begin{equation}
\overline{G}_i(s) = \frac{2 M_i}{(4 \pi)^2} \left(\left[ \frac{M_i^2-m_i^2}{s}-\frac{M_i-m_i}{M_i+m_i}\right] \log \frac{M_i}{m_i}+L_i(s) \right),\label{eq:defG}
\end{equation}
where
\begin{subequations}
\begin{align}
& s_{i-}=(m_i-M_i)^2,\\
& s_{i+}=(m_i+M_i)^2,
\end{align}
\end{subequations}
and for real $s$ and above threshold, $s > s_{i+}$
\begin{equation}
\label{eq:sourceSRS}
L_i(s+i\epsilon) = \frac{\lambda^{\frac{1}{2}}(s, m_i^2,M_i^2)}{s} \left(\log\left[\frac{1+\sqrt{\frac{s-s_{i+}}{s-s_{i-}}}}{1-\sqrt{\frac{s-s_{i+}}{s-s_{i-}}}} \right] - i \pi\right),
\end{equation}
and $\lambda(x,y,z)$ the ordinary K{\"a}llen function. 

The divergent contribution of the loop function, $G_i(s_{i+})$ in
Eq.~(\ref{eq:div}) needs to be renormalized.  We will
examine here two different renormalization schemes, widely used
in the literature.

On the one hand, we will  perform
one subtraction at certain scale $\sqrt{s}=\mu$, such that
\begin{equation}
G_i(\sqrt{s}=\mu) = 0\,.
\label{eq:subs}
\end{equation}
In this way,
\begin{equation}
G_i^\mu(s_{i+})  = - \overline{G}_i(\mu^2).
\end{equation}
so that
\begin{equation}
G_i^\mu(s) =\overline{G}_i(s) - \overline{G}_i(\mu^2).
\label{eq:relation}
\end{equation}
In addition,  we use the prescription adopted in
Ref.~\cite{Romanets:2012hm}, where $\mu$ is chosen to be  independent
of the total angular momentum $J$, common for all channels in a given
$CSI$ sector, and equal to
\begin{equation}
  \mu = \sqrt{\alpha \left(m_{th}^2+M_{th}^2 \right)}\label{eq:defmualpha} ,
\end{equation}
with $m_{th}$ and $M_{th}$ the masses of the meson and baryon of
the channel with the lowest threshold in the given $CSI$
sector~\cite{Hofmann:2005sw, Hofmann:2006qx},
and $\alpha$ a parameter that can be adjusted to
data~\cite{GarciaRecio:2008dp}. In what follows, we will refer to this
scheme as $\mu-RS$.
 
In the second RS, we make finite the UV divergent part of
the loop function using a sharp-cutoff regulator $\Lambda$ in momentum space,
which leads to~\cite{GarciaRecio:2010ki}
\begin{eqnarray}
G_i^\Lambda(s_{i+}) &=& \frac{1}{4\pi^2} \frac{M_i}{m_i+M_i} \left
(m_i\ln\frac{m_i}{\Lambda + \sqrt{\Lambda^2+m_i^2}}\right.\nonumber\\
&+& \left.  M_i\ln\frac{M_i}{\Lambda + \sqrt{\Lambda^2+M_i^2}} \right)\label{eq:uvcut} ,
\end{eqnarray}
and thus, for the UV cutoff case we have
\begin{equation}
G^{\Lambda}_i(s) =\overline{G}_i(s) + G_i^{\Lambda}(s_{i+}). \label{eq:uvcut2}
\end{equation}
Note that,  there are no cutoff effects in the  finite
$\overline{G}_i(s)-$loop function,  as it would happen if the two-body
propagator of Eq.~\eqref{eq:defG} would have been directly calculated using
the UV cutoff $\Lambda$.

If a common UV cutoff is employed for all channels within a
given $CSI$ sector, both RSs are independent and will lead to
different results. However, if one allows the freedom of using
channel-dependent cutoffs, the one-subtraction RS, $\mu-RS$,  is recovered by
choosing in each channel, $\Lambda_i$ such that
\begin{equation}
G^{\Lambda_i}_i(s_{i+})= -\overline{G}_i(\mu^2) .
\label{eq:subtraction}
\end{equation}

The dynamically-generated $\Omega_c$ resonances can be obtained as poles of the
scattering amplitudes in each $J$ sector for $(C=1,S=-2, I= 0)$.  We look at both
the first and second Riemann sheets (FRS and SRS) of the variable $\sqrt{s}$. The poles of the
scattering amplitude on the FRS that appear on the real
axis below threshold are interpreted as bound states. The poles that are found
on the SRS below the real axis and above threshold are
identified with resonances\footnote{Often we refer to all poles generically as
resonances, regardless of their concrete nature, since usually they can decay
through other channels not included in the model space.}.  The mass and the
width of the bound state/resonance can be found from the position of the pole
on the complex energy plane. Close to the pole, the $T$-matrix behaves
as 
\begin{equation}
T_{ij}(s) \simeq \frac{g_i g_j}{\sqrt{s}-\sqrt{s_R}}.
\end{equation}
The quantity $\sqrt{s_R}=M_R - \rm{i}\, \Gamma_R/2$ provides the mass ($M_R$) and the width
($\Gamma_R$) of the state, and $g_i$ is the complex coupling of the resonance to the channel $i$.

The couplings $g_i$ are obtained by first assigning an arbitrary sign
to one of them, say $g_1$. Then, we have that
\begin{equation}
g_1^2=\lim_{\sqrt{s}\rightarrow\sqrt{s_R}} (\sqrt{s}-\sqrt{s_R})T_{11}(s) ,
\end{equation}
and the other couplings result from
\begin{equation}
g_j = g_1 \lim_{\sqrt{s}\rightarrow\sqrt{s_R}} \frac{T_{1j}(s)}{T_{11}(s)} .
\end{equation}

In order to analyze the contribution of each baryon-meson channel to
the generation of a resonance, one has to not only analyze the
coupling but also the size of each baryon-meson loop, since the
product $g_i G_i(s_R)$ gives the strength of the wave function at the
origin for $s$-wave \cite{Gamermann:2009uq}.

\section{Results}
\label{sec:res}

The LHCb experiment has analyzed the $\Xi_c^+ K^-$ spectrum using $pp$
collisions and five new narrow excited $\Omega_c^0$ states have been
observed: the $\Omega_c^0(3000)$, $\Omega_c^0(3050)$,
$\Omega_c^0(3066)$, $\Omega_c^0(3090)$ and the $\Omega_c^0(3119)$, the
last three also seen in the $\Xi_c^{'+} K^-$ decay. Moreover, a sixth
broad structure around 3188 has also been found in the $\Xi_c^+ K^-$
spectrum.

\begin{figure}[t]
  \includegraphics[scale = 0.38]{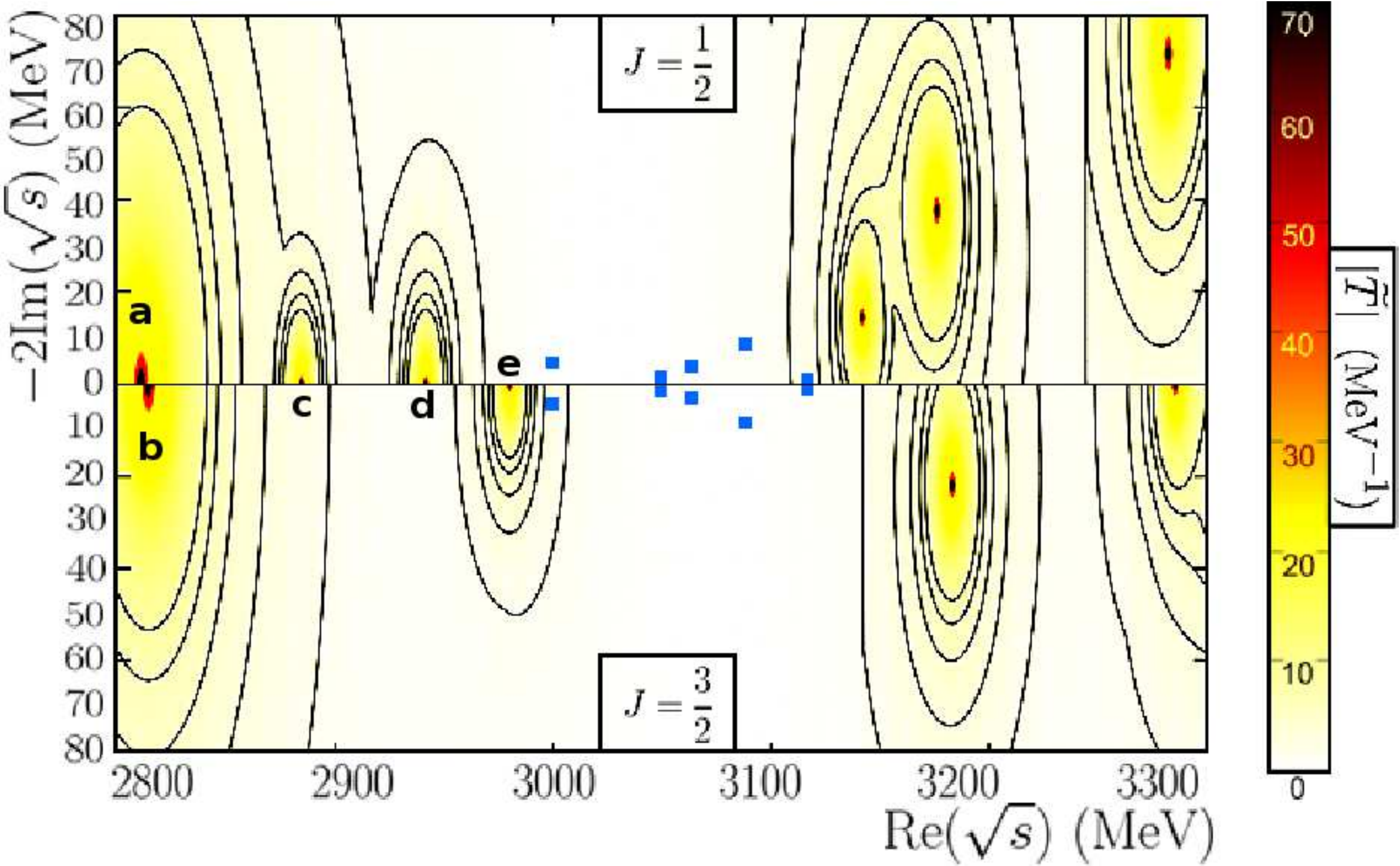}
  \caption{$\Omega_c (J=1/2)$  an $\Omega_c^* (J=3/2)$ odd-parity states, reported in
    Ref.~\cite{Romanets:2012hm}, coming from the most attractive ${\rm
      SU(6)}_{\rm lsf} \times$ HQSS representations. These five
    states, denoted as in Table~\ref{tab:table1}, are located below 3
    GeV for $J=1/2$ (upper plot) and $J=3/2$ (lower plot), while the
    five heavier resonant states above 3 GeV, also shown, come from less
    attractive  ${\rm SU(6)}_{\rm lsf}\times$HQSS multiplets, stemming from the exotic
    {\bf 4752} SU(8) representation. Since the dynamically generated
    states may couple differently to their baryon-meson components, we
    show the $ij-$channel independent quantity $|\tilde T(z)|_J = {\rm
      max}_j \sum_i |T_{ij}^J(z)|$, which allows us to identify all
    the resonances within a $J-$sector at once.  The blue dots
    correspond to the experimentally observed states. We display them
    both in the upper and lower plots because their spin is not
    determined.}
  \label{fig:alfa1}
\end{figure}

\begin{table}[t]
  \centering
  \caption{ $\Omega_c$ and $\Omega_c^*$ states obtained using $\alpha=1.16$}
  \label{tab:table2}
  \begin{tabular}{c|c|c|c||c|c}
      Name & $M_R$ (MeV) & $\Gamma_R$ (MeV) & $J$ & $M_R^{exp}$ & $\Gamma_R^{exp}$ \\
    \hline
    {\bf a} & 2922.2 & 0 & 1/2 & --- & --- \\
    \hline
    {\bf b} & 2928.1 & 0 & 3/2 & --- & --- \\
    \hline
   {\bf c} & 2941.3 & 0 & 1/2 & --- & --- \\
    \hline
    {\bf d}& 2999.9 & 0.06 & 1/2 & 3000.4 & 4.5 \\
    \hline
    {\bf e} & 3036.3 & 0 & 3/2 &  3050.2 & 0.8 \\
  \end{tabular}
\end{table}

\begin{figure*}[t]
\centering
  \includegraphics[scale = 0.1]{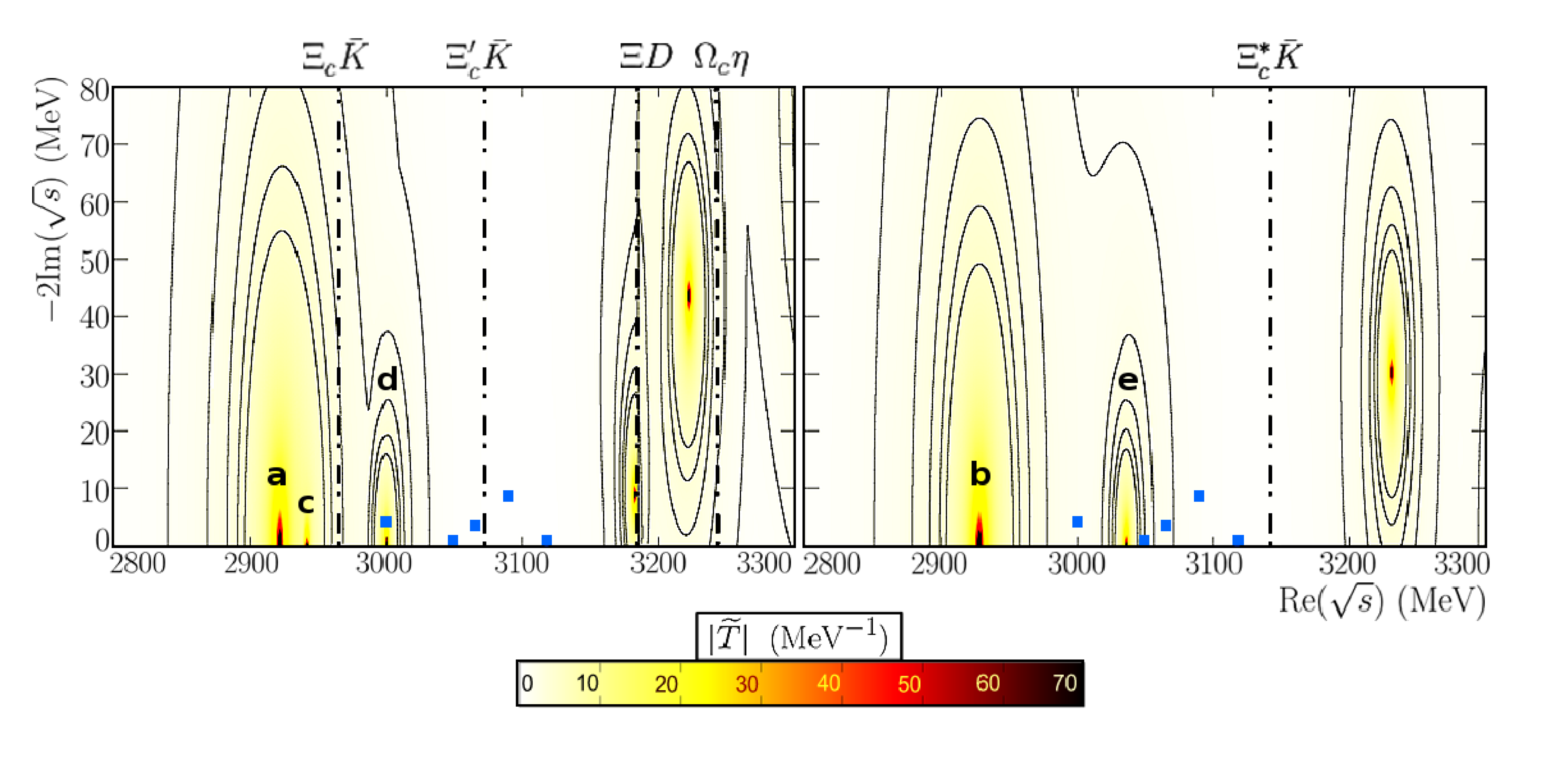}
  \caption{$\Omega_c$ and $\Omega_c^*$ states obtained within the
    scheme of Ref.~\cite{Romanets:2012hm} using $\alpha=1.16$. The
    left (right) plot shows the states dynamically generated for
    $J=\frac{1}{2}$ ($J=\frac{3}{2}$). The dotted blue points are the experimental
    observations, while some baryon-meson thresholds (dashed-dotted lines) are displayed for
    completeness. The function $|\tilde T(z)|_J$ is
    defined as in Fig.~\ref{fig:alfa1}. }
  \label{fig:alfa116}
\end{figure*}

\begin{table*}[t]
  \centering
  \caption{Properties of the $\Omega_c(2999.9)$ and
    $\Omega_c^*(3036.3)$ states, labeled as poles {\bf d} and {\bf e},
    respectively, obtained using $\alpha=1.16$. The first column
    displays the different baryon-meson channels coupled to
    $\Omega_c(2999.9)$, ordered by their threshold energies, in the
    $J=1/2$ sector. The second and third columns show the absolute
    value of the coupling and the product of the coupling times the 
    loop function at the pole position, respectively, for all
    baryon-meson coupled states. The fourth, fifth and sixth columns
    are equivalent to the first three columns but for
    $\Omega_c^*(3036.3)$ in the $J=3/2$ sector. }
  \label{tab:polesa}
  \begin{tabular}{l|c|r||l|c|r}
 $J=1/2$ & \multicolumn{2}{c||}{pole {\bf d}} &$J=3/2$
    & \multicolumn{2}{|c}{ pole {\bf e}} 
 \\ \hline

   \text{channel} & 	 $\ \ \ \ |g| \ \  \ \ $	& 	 $\ \  gG \ \text{\text{(MeV)}} \ \  \ \ $	&   \text{channel} &	$\ \ \ \ |g|\ \  \ \ $	& 	 $\ \  gG \ \text{(MeV)} \ \  \ \ $
 \\ \hline
$\Xi_c \bar{K}$	&$ 	0.1		$&$ 	-1.4	+	0.3	j	$&	$\Xi^*_c \bar{K}$	&$ 	1.9		   	$&$ 	-26.6		-0.1	j	   	$\\ \hline
$\Xi_c' \bar{K}$	&$ 	1.8		$&$ 	-27.1				$&	$\Omega_c^* \eta$	&$ 	1.7		   	$&$ 	16.3				   	$\\ \hline
$\Xi D$	&$ 	1.7		$&$ 	10.4				$&	$\Xi D^*$	&$ 	1.6		   	$&$ 	-8.5				   	$\\ \hline
$\Omega_c \eta$	&$ 	1.7		$&$ 	15.7				$&	$\Xi_c \bar{K}^*$	&$ 	1.6		   	$&$ 	-14				   	$\\ \hline
$\Xi D^*$	&$ 	0.8		$&$ 	-3.5		-0.1	j	$&	$\Xi^* D$	&$ 	0.5		   	$&$ 	-2.7				   	$\\ \hline
$\Xi_c \bar{K}^*$	&$ 	1.3		$&$ 	10.1				$&	$\Xi'_c \bar{K}^*$	&$ 	0.6		   	$&$ 	-4.9				   	$\\ \hline
$\Xi'_c \bar{K}^*$	&$ 	1.1		$&$ 	-7.3		-0.2	j	$&	$\Omega_c \omega$	&$ 	0		   	$&$ 	0.3					$\\ \hline
$\Omega_c \omega$	&$ 	0.1		$&$ 	0.7				$&	$\Xi^*_c \bar{K}^*$	&$ 	1.3		   	$&$ 	-8.9				   	$\\ \hline
$\Xi^*_c \bar{K}^*$	&$ 	0.6		$&$ 	3.6		-0.2	j	$&	$\Xi^* D^*$	&$ 	0.6		   	$&$ 	-2.4					$\\ \hline
$\Xi^* D^*$	&$ 	0.7		$&$ 	-2.6				$&	$\Omega_c^* \omega$	&$ 	0.1		   	$&$ 	0.4					$\\ \hline
$\Omega_c^* \omega$	&$ 	0		$&$ 	0				$&	$\Omega D_s$	&$ 	0.8		   	$&$ 	-3.3				   	$\\ \hline
$\Omega_c \eta'$	&$ 	0.5		$&$ 	2.5				$&	$\Omega_c \phi$	&$ 	0.6		   	$&$ 	3.5				   	$\\ \hline
$\Omega_c \phi$	&$ 	1.1		$&$ 	5.4	+	0.1	j	$&	$\Omega_c^* \eta'$	&$ 	0.5		   	$&$ 	2.8				   	$\\ \hline
$\Omega D_s^*$	&$ 	1.2		$&$ 	-3.7				$&	$\Omega D_s^*$	&$ 	1		   	$&$ 	-3.4				   	$\\ \hline
$\Omega_c^*  \phi$	&$ 	0.6		$&$ 	-2.9	+	0.1	j	$&	$\Omega_c^* \phi$	&$ 	1.2		   	$&$ 	6.5				   	$\\ \hline

  \end{tabular}
\end{table*}

As mentioned, the unitarized coupled-channel model of
Ref.~\cite{Romanets:2012hm}, based on a ${\rm SU(6)}_{\rm
  lsf}\times$HQSS- extended WT interaction, predicted five excited
odd-parity $\Omega_c$ states with spins $1/2$ and $3/2$ and masses
below 3 GeV (Table~\ref{tab:table1}). In Fig.~\ref{fig:alfa1}, the
positions of the three $\Omega_c$ states (upper panel) and the two
$\Omega_c^*$ (lower panel) are shown.  We see that all masses are
below 2.98 GeV, which makes difficult to identify any of them with any
of the LHCb resonances. Masses and widths of other five
resonances above 3 GeV are also displayed in Fig.~\ref{fig:alfa1}.
These resonances were not discussed in Ref.~\cite{Romanets:2012hm},
and are much more uncertain,  as
they result from less attractive ${\rm SU(6)}_{\rm lsf}\times$HQSS
multiplets related to the exotic {\bf 4752} SU(8) irreducible
representation.

All these states have been dynamically generated by solving a
coupled-channel BSE using a ${\rm SU(6)}_{\rm lsf}\times$HQSS-extended
WT interaction as a kernel (see Sec.~\ref{formalism}). The
baryon-meson loops have been renormalized implementing
one-substraction at the scale $\mu = \sqrt{\alpha \left(m_{th}^2+M_{th}^2 \right)}$,
with $\alpha=1$. This RS was chosen following the works of
Refs.~\cite{Hofmann:2005sw,Hofmann:2006qx}, where it was claimed that
such a choice guarantees an approximate crossing symmetry. Moreover it
also allowed for a successfully description of the
$\Lambda_c(2595)$ and $\Lambda_c(2625)$ resonances, with
almost\footnote{Only, the $\alpha$ parameter in
  Eq.~\eqref{eq:defmualpha} was slightly reduced from the default
  value of 1 advocated in Ref.~\cite{Hofmann:2005sw, Hofmann:2006qx}.}
no-free parameters~\cite{GarciaRecio:2008dp}.

However, it is possible to allow for some freedom
and slightly modify the choice of the subtraction point by changing
the value of $\alpha$. In this way, we might move up in energy the
states found in Ref.~\cite{Romanets:2012hm} and compiled in
Table~\ref{tab:table1}, and try to identify some of them with the
experimentally observed $\Omega_c^{(*)}$ states.  We concentrate our study on
those states as they are the ones most likely to exist since they
originate from the most attractive ${\rm SU(6)}_{\rm lsf} \times$ HQSS
representations.

Masses become higher when $\alpha$ becomes greater than one. Allowing
for just moderately changes, we find that for $\alpha=1.16$ the two
last states, labeled with {\bf d} and {\bf e} in
Table~\ref{tab:table1}, are now located near the experimental
$\Omega_c(3000)$ and $\Omega_c(3050)$, with masses 2999.9 MeV and
3036.3 MeV, respectively, while their widths are almost zero. The
poles found with this new value of $\alpha$ are compiled in Table
\ref{tab:table2} and displayed in Fig.~\ref{fig:alfa116}.  Moreover,
the analysis of the product of the coupling times the loop function at
the pole, $g_i G_i(s_R)$, of Table \ref{tab:polesa} allows us to study
the importance of the different baryon-meson channels to the dynamical
generation of the $\Omega_c$ and $\Omega_c^*$ states. In particular,
the state at 2999.9 MeV is mainly a $\Xi_c^{'+} \bar K$ molecular
state that also couples strongly to $\Omega_c \eta$, $\Xi D$ and
$\Xi_c \bar K^*$.  As for the state at 3036.3 MeV, the dominant
$\Xi_c^* \bar K$ channel can be reconciled with the experimentally
seen decay $\Xi_c^+ K^-$, if one allows for the $\Xi_c^* \bar K \to
\Xi_c \bar K$ $d-$wave transition, that does not involve the exchange of the
charm-quark.

In view of the previous results, we explore a different RS to evaluate
the impact of the renormalization procedure in the predictions of the
$\Omega_c$ and $\Omega_c^*$ low-lying odd parity states, aiming at
providing an alternative description for some of the states observed
by LHCb. Thus, we allow for a variation of the subtraction constants
in each channel different to that imposed within the $\mu-RS$, but
still in a controlled way. For that purpose, we use the relation
between the subtraction constants and the cutoff scheme given in
Eqs.~(\ref{eq:uvcut}) and (\ref{eq:uvcut2}), and employ a common UV
cutoff for all baryon-meson loops within reasonable limits. In this
way, on the one hand, we avoid any fictitious reduction of any
baryon-meson channel by using a small value of the cutoff and, on the
other hand, we prevent an arbitrary variation of the subtraction
constants\footnote{This will induce an enormous freedom difficult to fix
with the scarce available data.}, since we correlate all of them to a
reasonable value of the UV cutoff, while still keeping the full
analyticity of the baryon-meson loops, as discussed below
Eq.~(\ref{eq:uvcut2}).

\begin{table}[h]
  \caption{ $\Omega_c$ and $\Omega_c^*$ states calculated using the
    subtraction constants associated to a cutoff of $\Lambda=1090$
    MeV. We identify experimentally two $J=1/2$ and one $J=3/2$ states.}
  \label{tab:table3}
  \begin{tabular}{c|c|c|c||c|c}
      Name & $M_R$ (MeV) & $\Gamma_R$ (MeV) & $J$ & $M_R^{exp}$ & $\Gamma_R^{exp}$ \\
    \hline
    {\bf a} & 2963.95 & 0.0 & 1/2 & --- & --- \\
    \hline
    {\bf c} & 2994.26 & 1.85 & 1/2 & 3000.4 & 4.5 \\
    \hline
    {\bf b} & 3048.7 & 0.0 & 3/2 & 3050.2  &  0.8 \\
    \hline
    {\bf d} & 3116.81 & 3.72 & 1/2 & 3119.1/ 3090.2  & 1.1/ 8.7   \\
    \hline
    {\bf e} & 3155.37 & 0.17 & 3/2 &  --- & --- \\
  \end{tabular}
\end{table}

To identify our five dynamically generated
$\Omega_c$ and $\Omega_c^*$ states of Table~\ref{tab:table1} using the new
subtraction constants, we first need to determine how the masses (and
widths) of our generated states change as we adiabatically vary the
values of the subtraction constants. This can be done by
\begin{equation}
G_i(s) = \overline{G}_i(s)-(1-x) \overline{G}_i(\mu^2)+x G_i^{\Lambda}(s_{i+}),
\end{equation}
where $x$ is a parameter that changes slowly from $0$ to $1$, and
$\mu^2=(m_{th}^2+M_{th}^2)$.  In this manner, we can follow in the
complex energy plane the original $\Omega_c$ and $\Omega_c^*$ as we
modified our prescription to use a common cutoff for the computation
of the subtraction constants.

Our results for the $\Omega_c$ and $\Omega_c^*$ are shown in
Table~\ref{tab:table3} for a fixed cutoff of $\Lambda=1090$ MeV. In
this case, we find that three poles (those previously named {\bf c},
{\bf b} and {\bf d}) can be identified with the three experimental
states at 3000 MeV, 3050 MeV and 3119 or 3090 MeV.  The identification is
possible not only due to the closeness in energy to the experimental
ones but also because of the dominant contribution of the experimental
$\Xi_c \bar K$ and $\Xi_c^{'} \bar K$ channels to their dynamical
generation. The contribution is measured by the product $gG$ at the
pole, as reported in Table~\ref{tab:poles1} for $J=1/2$ and
Table~\ref{tab:poles2} for $J=3/2$. For the $J=1/2$ state at 2994 MeV
(pole {\bf c}), we observe a significant contribution of the
$\Xi_c^{'} \bar K$ and $\Xi_c \bar K$ channels, while $\Omega_c \eta$
is also relevant.  We identify this state with $\Omega_c(3000)$.  As
for the $J=1/2$ state at 3117 MeV (pole {\bf d}), the dominant
contribution comes from $\Xi D$ but also from $\Xi_c \bar K^*$, $\Xi
D^*$ and $\Xi_c \bar K$. Thus, we can identify this state with
$\Omega_c(3119)$ or the $\Omega_c(3090)$ given its proximity in
mass. Moreover, a sizable width of $8.7 \pm 1.0 \pm 0.8$ MeV is
reported for the latter state in Ref.~\cite{Aaij:2017nav} to be compared
with the one around 4 MeV found here for the state {\bf d}.  Finally, the
$J=3/2$ state at 3049 MeV (pole {\bf b}) could be identified with $\Omega_c
(3050)$ as it couples strongly to $\Xi_c^* \bar K$ and $\Xi_c\bar
K^*$, channels connected to $\Xi_c \bar K$ by $d-$wave transitions, while
having also an important contribution from $\Omega_c^* \eta$. In
summary, two $J=1/2$ and one $J=3/2$ can be identified experimentally
for a cutoff of $\Lambda=1090$ MeV.

\begin{figure*}[ht!]
  \includegraphics[scale = 0.4]{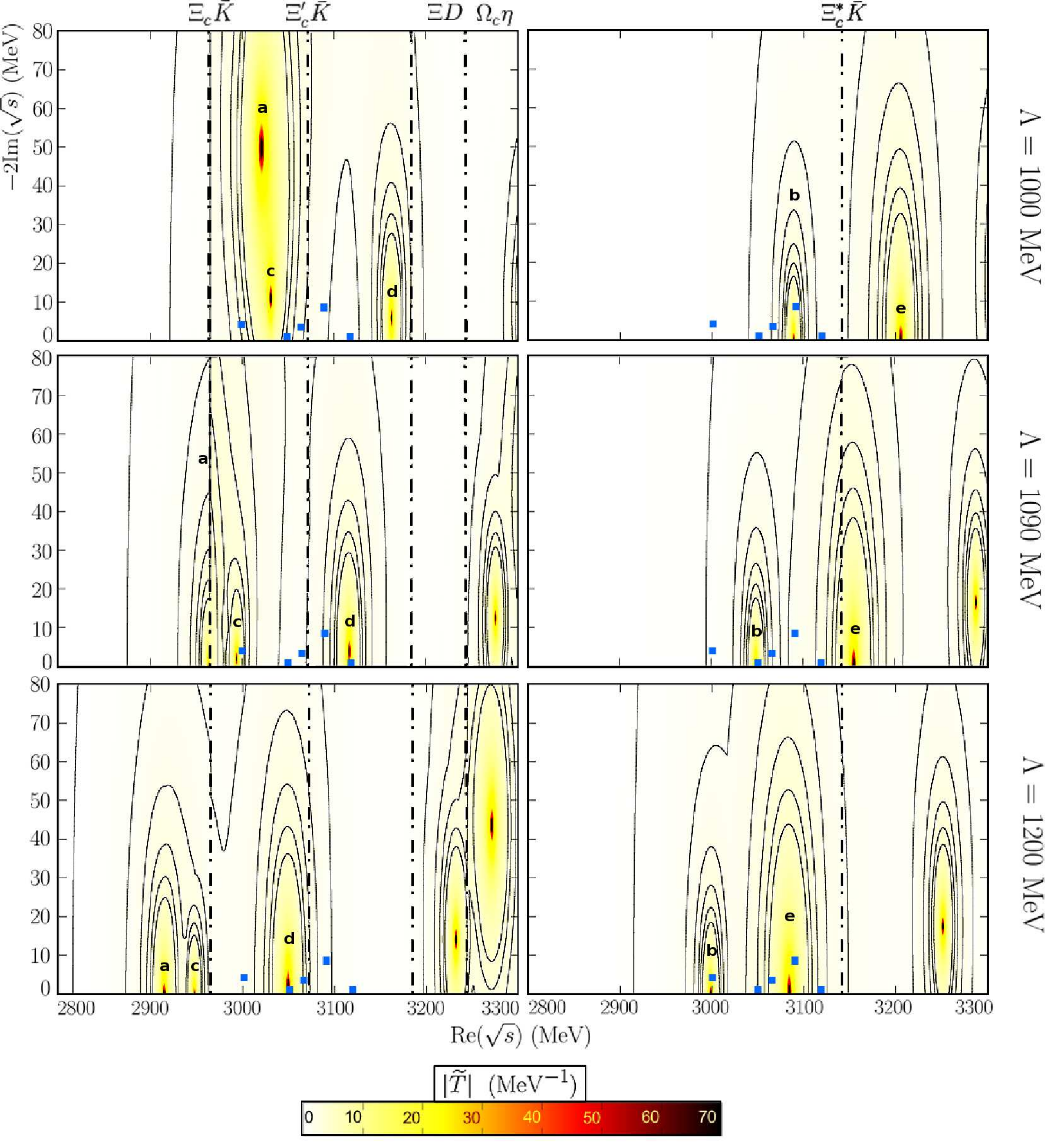}
  \caption{$\Omega_c$ and $\Omega_c^*$ states for different UV
    cutoffs.  The blue squares indicate the experimental
    points. Dashed-dotted lines represent the closest baryon-meson
    thresholds. The left plots are for $J=\frac{1}{2}$ and the right
    ones for $J=\frac{3}{2}$, while the function $|\tilde T(z)|_J$ is
    defined as in Fig.~\ref{fig:alfa1}. For the two largest values of
    $\Lambda$, some resonant states from less attractive ${\rm
      SU(6)}_{\rm lsf}\times$HQSS multiplets, stemming from the exotic {\bf 4752} SU(8)
    representation, are also visible  in the region of  higher masses.}
  \label{fig:lambdas}
\end{figure*}

In order to assess the dependence of our results on the cutoff, we
have examined lower and higher values. As indicated before, the variation in the cutoff scale changes the value of the subtraction constant. This variation is related to the change of the size of higher order corrections in the meson-baryon scattering amplitude that are not known and not fixed by unitarization. Below 800 MeV, all resonances
become heavier and much wider than the observed LHCb states.
Actually, a clear identification between our results and some of the
experimental states is not possible until a value of $\Lambda \sim
1000$ MeV. For cutoffs bigger than 1300--1350 MeV, the $\Omega_c$ and
$\Omega_c^*$ states coming from the most attractive ${\rm SU(6)}_{\rm
  lsf} \times$ HQSS representations appear well below 3 GeV, and we
can neither make an identification between those states and the LHCb
spectrum. In Fig.~\ref{fig:lambdas}, we show the obtained pole
positions for $\Lambda= 1090$ MeV (Table~\ref{tab:table3}) and two
additional cutoffs, around 100 MeV smaller and bigger, respectively,
than this central one.  It can be seen that for $\Lambda=$1090 MeV and
$\Lambda=$1200 MeV, a maximum number of three states can be
identified. As compared to the $\Lambda=$1090 MeV case previously
discussed, for $\Lambda=$1200 MeV we can identify two $\Omega_c^*$
states with $J=3/2$ at 3000 MeV and 3090 MeV, whereas a $J=1/2$
$\Omega_c$ is seen at 3050 MeV. The $J=1/2$ state at 3050 MeV
corresponds now to the {\bf d} state, that for $\Lambda=1090$ MeV was
identified with the $\Omega_c(3119)$ or $\Omega_c(3090)$
resonances, and it has a dominant
$ \Xi D$ component. It might still be the $\Omega_c(3090)$.
The $J=1/2$ {\bf c} pole now moves well below 3
GeV and this makes difficult its identification with any of the LHCb
states.   In the $J=3/2$ sector, the resonance that appears a 3000 MeV is
the pole {\bf b} and strongly couples to $ \Xi_c^* \bar K$ and
$\Xi_c\bar K^*$, as already mentioned above.  The additional $J=3/2$
state at 3090 MeV is the pole {\bf e} in the nomenclature used in
Table~\ref{tab:table3} for $\Lambda=1090$ MeV, and as it can be seen
there, it has a large $\Xi D^* $ molecular component, and it could be
associated to the $\Omega_c(3119)$ or $\Omega_c(3090)$ LHCb
resonances. In all three
cases and in order to make the experimental identification possible, a
significant coupling to the $\Xi_c \bar K$ channel could be obtained,
often via $\Xi_c^* \bar K$ and $\Xi_c \bar K^*$ allowing for the
$d-$wave transitions. In summary we see that by changing the UV cutoff,  the
pole positions of the dynamically generated states are modified making
more plausible different identifications between some of these states and those
observed by LHCb.

As mentioned in the Introduction, the molecular nature of the five
$\Omega_c$ narrow states has been recently analyzed in
Refs.~\cite{Montana:2017kjw,Debastiani:2017ewu} as well as the
observed broad structure around 3188 MeV in
Ref.~\cite{Wang:2017smo}. In Ref.~\cite{Montana:2017kjw} the
interaction of the low-lying mesons (pseudoscalar and vector mesons
separately) with the ground-state $1/2^+$ baryons in the $C=+1$, $S-2$
and $I=0$ sector has been built from $t$-channel vector meson
exchanges. Two $J=1/2$ baryon-meson molecular states could be
identified with the experimental $\Omega_c(3050)$ and
$\Omega_c(3090)$, mostly having the state at 3050 MeV a $ \Xi_c^{'}
\bar K$ component with an admixture of $ \Omega_c \eta$, while the
3090 MeV would be a $ \Xi D$ molecule. These results have been
reproduced in the $J=1/2$ sector in Ref.~\cite{Debastiani:2017ewu},
within a local hidden gauge approach extended to the charm sector that
also incorporates baryon $3/2^+$-pseudoscalar meson components. This
is because the diagonal terms in the interaction kernel are the same
in both models and these two $\Omega_c$ states do not couple to baryon
$1/2^+$-vector meson channels in
Refs.~\cite{Montana:2017kjw,Debastiani:2017ewu}. Furthermore, by
incorporating baryon $3/2^+$-pseudoscalar meson states, a $J=3/2$
baryon-meson molecular state has been also identified in
Ref.~\cite{Debastiani:2017ewu} with the experimental
$\Omega_c(3119)$. This state would be a baryon $3/2^+$- pseudoscalar
meson molecule with large couplings to $ \bar K \Xi_c^*$ and
$\Omega^*_c\eta$.

In this work and for $\Lambda=1090$ MeV, we have also obtained
three baryon-meson molecular states that couple predominantly to $\bar
K \Xi_c^{'}$, $D \Xi$ and $\bar K \Xi_c^*$, respectively, but with a
different experimental assignment of masses, that is, $J=1/2$
$\Omega_c(3000)$ and $J=1/2$ $\Omega_c(3119)$ or $\Omega_c(3090)$, and $J=3/2$
$\Omega_c(3050)$, which correspond to poles {\bf c} and {\bf d}, and
{\bf b}, respectively. However, the $g_i G_i(s_R)$  strengths for the dominant
channels found in this work are in reasonable good agreement with those
given in Ref.~\cite{Debastiani:2017ewu}. As we have illustrated in
Fig. ~\ref{fig:lambdas}, our predictions for masses are subjected to sizeable
uncertainties,  which might lead to confusions in the assignments to the
LHCb states proposed in this work. 

Nevertheless we should highlight that, we use here a different
regularization scheme of the loop functions and different interaction
matrices than in the works of
Refs.~\cite{Montana:2017kjw,Debastiani:2017ewu} that should explain
the differences found. Note that the matrix elements involving the
interaction of Goldstone-bosons and heavy-baryons are fixed by chiral
symmetry and should agree in the three approaches. The differences
come from channels involving $D$, $D^*$ and light-vector mesons, where
HQSS does not completely fix the interactions. Furthermore, in the
models of Refs.~\cite{Montana:2017kjw,Debastiani:2017ewu} some HQSS
breaking terms suppressed by the heavy-quark-mass are accepted.  In
addition, we incorporate the mixing of channels involving pseudoscalar
mesons with channels involving vector mesons, while such mixings are
claimed to be negligible in the case of
Ref.~\cite{Debastiani:2017ewu}.  Our model also incorporates the
contribution of baryon-meson states of higher mass than those included
in Refs.~\cite{Montana:2017kjw,Debastiani:2017ewu}, though, those
heavier baryon-meson channels do not give any relevant contribution to
the generation of the low-lying $\Omega_c$ and $\Omega_c^*$ states.

In Ref.~\cite{Wang:2017smo} the broad structure observed by the LHCb
Collaboration around 3188 MeV has been analysed as the superposition
of two $D \Xi$ bound states within the Bethe-Salpeter formalism in the
ladder and instantaneous approximation. As can be seen in
Fig.~\ref{fig:lambdas}. we also generate resonances in this
region, but it is difficult to reach any conclusion since most likely,
we would have to consider also some states from less attractive ${\rm
  SU(6)}_{\rm lsf}\times$HQSS multiplets, stemming from the exotic {\bf 4752} SU(8)
representation~\cite{Romanets:2012hm}. A candidate of a loosely
bound molecular state with a large  $\Xi^*_c\bar K$ component and a
mass around 3140 MeV is  also predicted in Ref.~\cite{Chen:2017xat}. It
results from  $\Xi^*_c\bar K/\Xi_c\bar K^*/\Xi_c^\prime\bar K^* $
coupled-channel dynamics using a one-boson-exchange potential. It is
difficult to associate such state with any of the predictions
obtained here from the scheme of Ref.~\cite{Romanets:2012hm}, since the
work of Ref.~\cite{Chen:2017xat} does not consider $\Xi^{(*)}D^{(*)}$
channels.

\section{Conclusions}

We have reviewed the RS used in the unitarized coupled-channel model
of Ref.~\cite{Romanets:2012hm} and its impact in the $C=1$, $S=-2$,
and $I=0$ sector, where five $\Omega_c$ states have been recently
observed by the LHCb Collaboration~\cite{Aaij:2017nav}.  A
coupled-channel BSE, with a ${\rm SU(6)}_{\rm lsf}\times$HQSS-extended
WT meson-baryon interaction, is solved in \cite{Romanets:2012hm}
within the on--shell approximation, and adopting a one-subtraction RS
at fixed scale for all channels, as advocated in
Refs.~\cite{Hofmann:2005sw, Hofmann:2006qx}.  Five odd-parity
$\Omega_c, \Omega^*_c$ states, coming from the most attractive ${\rm
  SU(6)}_{\rm lsf}\times$HQSS representations, are dynamically
generated, but with masses below 2.98 GeV that cannot be easily
identified with any of the LHCb resonances, located all of them above
3 GeV. Predicted masses can be moved up by implementing a different
RS. We have explored two different scenarios, introducing at most only
one additional undetermined parameter in the scheme. In the first one,
the common energy-scale used in \cite{Romanets:2012hm} to perform the
subtractions is modified allowing for moderate variations.  In the
second one, a common UV cutoff is used to render finite the UV
divergent loop functions in all channels. In both cases, we could move
two or three states in the region between 3 and 3.1 GeV, where the
LHCb resonances lie. In particular, when we use $\Lambda=1090$ MeV, we
obtain three baryon-meson molecular states (poles {\bf c} and {\bf d},
and {\bf b}) that couple predominantly to $\bar K \Xi_c^{'}$, $D \Xi$
and $\bar K \Xi_c^*$, and can be easily related to the LHCb resonances
and to results of Refs.~\cite{Montana:2017kjw,Debastiani:2017ewu}.
Thus for the dominant channels, we obtain strengths for the wave
function at the origin in a reasonable good agreement with those found
in Ref.~\cite{Debastiani:2017ewu}.  There exist, however, some
disagreements in the predictions for the masses, which need to be
taken with some caution. At least, our predictions for masses are
subjected to sizable uncertainties, which might lead also to
confusions in the assignments to the LHCb states proposed in this
work. 

In summary, we can conclude that some (probably at least three) of the states observed by
LHCb~\cite{Aaij:2017nav} will have odd parity and spins $J=1/2$ and
$J=3/2$. Moreover, those associated to the poles {\bf b} with $J=3/2$ and
{\bf c} with $J=1/2$ would belong to the same  ${\rm SU(6)}_{\rm lsf}$
$\times$ HQSS multiplets~\cite{Romanets:2012hm,GarciaRecio:2012db} that the strangeness-less $\Lambda_c(2595)$ and
$\Lambda_c(2625)$, and $\Lambda_b(5912)$ and
$\Lambda_b(5920)$ resonances in the charm and bottom sectors, respectively.

\label{conc}

\section{Acknowledgements}
The authors warmly thank V. R. Debastiani, E. Oset and A. Ramos for valuable discussions. L.T. acknowledges support from the Heisenberg Programme of the
Deutsche Forschungsgemeinschaft under the Project Nr. 383452331, the Ram\'on y Cajal research
programme and THOR COST Action CA15213. R. P. Pavao wishes to thank
the Generalitat Valenciana in the program Santiago Grisolia. This
research is supported by the Spanish Ministerio de Econom\'ia y
Competitividad and the European Regional Development Fund, under
contracts FIS2014-51948-C2-1-P, FIS2017-84038-C2-1-P, FPA2013-43425-P,
FPA2016-81114-P and SEV-2014-0398 and by Generalitat Valenciana under
contract PROMETEOII/2014/0068.

\begin{table*}[ht]

  \centering

  \caption{$J=1/2$ $\Omega_c$ states, labeled as poles {\bf a}, {\bf
      c} and {\bf d}, calculated using the subtraction constants
    determined by a unique UV cutoff $\Lambda=1090$ MeV (see
    Eq.~(\ref{eq:uvcut})). The first column displays the different
    baryon-meson coupled channels, ordered by their threshold
    energies. The subsequent columns show the absolute value of the
    coupling and the product of the coupling times the loop function
    at the pole for all baryon-meson coupled states for pole {\bf a}
    at 2963.95 MeV (second and third columns), pole {\bf c} at 2994.26 MeV
    (fourth and fifth columns) and pole {\bf d} at 3116.81 MeV (sixth and
    seventh columns). Poles {\bf c} at 2994.26 MeV and {\bf d} at 3116.81 MeV
    might be identified with the experimental $\Omega_c(3000)$ and
    the $\Omega_c(3119)$ or $\Omega_c(3090)$, respectively.}

  \label{tab:poles1}

  \begin{tabular}{l||c|r||c|r||c|r}

  & \multicolumn{2}{c||}{pole {\bf a}} & \multicolumn{2}{|c||}{pole
      {\bf c}} & \multicolumn{2}{|c}{pole {\bf d}}

 \\ \hline

   channel	& 	 $\ \ \ \ \ |g| \ \  \ \ $	&
   $\ \  gG \ \text{(MeV)}  \ \  \ \ $	& 	
$\ \ \ \ |g| \ \  \ \ $	& 	 $\ \  gG \ \text{(MeV)} \ \  \ \ $	&$\ \ \ \ |g| \ \  \ \ $	& 	 $\ \ gG \ \text{(MeV)} \ \  \ \ $ \\

       \hline
$\Xi_c \bar{K}$	&$ 	0.9		$&$ 	-33.0		-0.1	j	$&$	0.3		$&$ 	-10.2	+	6.0	j	$&$	0.3		$&$ 	-11.7	+	2.2	j		$\\ \hline
$\Xi_c' \bar{K}$	&$ 	0.4		$&$ 	-7.3				$&$	1.7		$&$ 	39.1	+	0.9	j	$&$	0.0		$&$ 	-0.6	+	0.1	j		$\\ \hline
$\Xi D$	&$ 	1.8		$&$ 	10.1				$&$	1.0		$&$ 	-6.4		-2.1	j	$&$	2.3		$&$ 	-26.9		-1.1	j		$\\ \hline
$\Omega_c \eta$	&$ 	0.4		$&$ 	4.1				$&$	1.9		$&$ 	-22.7		-0.5	j	$&$	0.3		$&$ 	-4.6					$\\ \hline
$\Xi D^*$	&$ 	1.7		$&$ 	3.6				$&$	1.4		$&$ 	3.5		-0.9	j	$&$	2.2		$&$ 	12.5		-0.8	j		$\\ \hline
$\Xi_c \bar{K}^*$	&$ 	0.0		$&$ 	-0.1				$&$	1.8		$&$ 	-8.7	+	0.2	j	$&$	1.8		$&$ 	17.4	+	0.1	j		$\\ \hline
$\Xi'_c \bar{K}^*$	&$ 	0.9		$&$ 	0.4				$&$	1.4		$&$ 	1.8		-0.3	j	$&$	0.2		$&$ 	-0.7		-0.6	j		$\\ \hline
$\Omega_c \omega$	&$ 	0.5		$&$ 	-0.4				$&$	0.6		$&$ 	-1.0	+	0.2	j	$&$	0.3		$&$ 	1.7	+	0.1	j		$\\ \hline
$\Xi^*_c \bar{K}^*$	&$ 	1.2		$&$ 	-2.0				$&$	0.3		$&$ 	0.1	+	0.2	j	$&$	1.5		$&$ 	3.8		-0.4	j		$\\ \hline
$\Xi^* D^*$	&$ 	0.2		$&$ 	0.4				$&$	0.9		$&$ 	1.7		-0.1	j	$&$	2.5		$&$ 	0.4		-0.1	j		$\\ \hline
$\Omega_c^* \omega$	&$ 	0.4		$&$ 	0.5				$&$	0.1		$&$ 	0.0	+			$&$	0.9		$&$ 	-2.7	+	0.1	j		$\\ \hline
$\Omega_c \eta'$	&$ 	0.1		$&$ 	-0.6				$&$	0.2		$&$ 	1.0	+	0.1	j	$&$	0.6		$&$ 	0.8					$\\ \hline
$\Omega_c \phi$	&$ 	0.4		$&$ 	2.6				$&$	1.1		$&$ 	7.2		-0.6	j	$&$	0.1		$&$ 	0.2		-0.3	j		$\\ \hline
$\Omega D_s^*$	&$ 	0.3		$&$ 	2.0				$&$	0.1		$&$ 	-0.8		-0.4	j	$&$	1.9		$&$ 	-9.2		-0.2	j		$\\ \hline
$\Omega_c^*  \phi$	&$ 	0.8		$&$ 	6.5				$&$	0.4		$&$ 	-2.8		-1.2	j	$&$	0.6		$&$ 	3.4		-0.5	j		$\\ \hline

  \end{tabular}
\end{table*}

\begin{table*}[ht]

  \centering

  \caption{$J=3/2$ $\Omega_c^*$ states, labeled as poles {\bf b} and
    {\bf e},
    calculated using the subtraction constants determined by a unique
    UV cutoff $\Lambda=1090$ MeV (see  Eq.~(\ref{eq:uvcut})).  The first column displays the
    different baryon-meson coupled channels, ordered by their
    threshold energies, for $J=3/2$. The subsequent columns show
    the absolute value of the coupling and the product of the coupling
    with the loop function at the pole for all baryon-meson
    coupled states for pole {\bf b} at 3048.7 MeV (second and third columns)
    and pole {\bf e} at 3155.37 MeV (fourth and fifth columns). Pole
    {\bf b} at
    3048.7 MeV might be identified with the experimental
    $\Omega_c(3050)$.}

  \label{tab:poles2}

  \begin{tabular}{l||c|r||c|r}

  & \multicolumn{2}{c||}{pole {\bf b}} & \multicolumn{2}{|c}{pole {\bf
        e}}

 \\ \hline

   channel	& 	 $\ \ \ \ \ |g| \ \  \ \ $	&
   $\ \  gG \ \text{(MeV)}  \ \  \ \ $	& 	
$\ \ \ \ |g| \ \  \ \ $	& 	 $\ \  gG \ \text{(MeV)} \ \  \ \ $ \\

       \hline
$\Xi^*_c \bar{K}$	&$ 	1.8		$&$ 	-38.8		-0.1	j	$&$	0.1		$&$ 	-4.3	+	0.1	j	$\\ \hline
$\Omega_c^* \eta$	&$ 	1.8		$&$ 	20.1				$&$	0.8		$&$ 	13.3		-0.3	j	$\\ \hline
$\Xi D^*$	&$ 	0.8		$&$ 	-3.0				$&$	3.6		$&$ 	-24.4				$\\ \hline
$\Xi_c \bar{K}^*$	&$ 	2.1		$&$ 	-14.0				$&$	0.9		$&$ 	10.5	+	0.2	j	$\\ \hline
$\Xi^* D$	&$ 	0.9		$&$ 	1.9				$&$	2.2		$&$ 	-10.7				$\\ \hline
$\Xi'_c \bar{K}^*$	&$ 	0.5		$&$ 	-1.3				$&$	0.1		$&$ 	-0.6	+	0.1	j	$\\ \hline
$\Omega_c \omega$	&$ 	0.3		$&$ 	1.0				$&$	0.4		$&$ 	-2.9				$\\ \hline
$\Xi^*_c \bar{K}^*$	&$ 	1.2		$&$ 	-0.7				$&$	0.6		$&$ 	2.4	+	0.1	j	$\\ \hline
$\Xi^* D^*$	&$ 	1.1		$&$ 	-1.2				$&$	2.4		$&$ 	-2.3				$\\ \hline
$\Omega_c^* \omega$	&$ 	0.4		$&$ 	0.4				$&$	0.2		$&$ 	-1.0				$\\ \hline
$\Omega D_s$	&$ 	0.1		$&$ 	-0.4				$&$	1.4		$&$ 	2.1				$\\ \hline
$\Omega_c \phi$	&$ 	0.5		$&$ 	-2.6				$&$	0.2		$&$ 	-0.4				$\\ \hline
$\Omega_c^* \eta'$	&$ 	0.1		$&$ 	-0.5				$&$	0.8		$&$ 	-2.0				$\\ \hline
$\Omega D_s^*$	&$ 	0.2		$&$ 	-1.1				$&$	1.9		$&$ 	8.1				$\\ \hline
$\Omega_c^* \phi$	&$ 	1.1		$&$ 	-7.6				$&$	0.1		$&$ 	0.4				$\\ \hline
  \end{tabular}
\end{table*}

\bibliographystyle{plain}

\end{document}